# Boundary $\sigma$-model and corrections to D-brane actions

Alexander Barabanschikov

*Department of Physics, Northeastern University*
*Boston, MA 02115, USA*

email: abaraban@lynx.neu.edu

**Abstract.**

We consider a $\sigma$-model formulation of open string theory in the presence of D-branes. We perform two-loop computations and discuss gravitational corrections to Born-Infeld action when branes are non-trivially embedded in a curved ambient space. In particular for the case of a stack of $N$ coincident D-branes we analyze couplings of the form $R_{ijkl}[\Phi^i\Phi^j][\Phi^k\Phi^l]$.

# 1 Introduction

To understand the dynamics of D-branes it is very important to study the low energy effective action. For bosonic strings to the leading order in $\alpha'$ it is given by the Dirac-Born-Infeld (DBI) action [1, 2]. For superstrings there is additional Wess-Zumino term describing the coupling of a brane to Ramond-Ramond fields [3][1]. In this paper we discuss certain higher order corrections to DBI action depending on the embedding of branes in a curved ambient space.

Suppose we have a Dp-brane non-trivially embedded in a target space. In Einstein frame the effective action for a single D-brane is given by

$$S_{DBI} = -\tau_p \int d\sigma^{p+1} e^{\Phi(-1-\gamma(p+1)/2)} \sqrt{-det(\tilde{G}_{\alpha\beta} + e^{\gamma\Phi}(\tilde{B}_{\alpha\beta} + 2\pi\alpha' F_{\alpha\beta}))} \quad (1)$$

Here $\gamma = -\frac{4}{D-2}$. We are using Greek letters $(\mu, \nu, ...)$ for space-time coordinates, $(\alpha, \beta, ...)$ for coordinates on the brane and Latin letters (i,j,...) for coordinates transverse to the brane. Thus in the case of a p-brane embedded in a D-dimensional target space: $\mu = 0,...,D-1$, $\alpha=0,...,p$ and i=p+1,...,D-1. $\{\sigma^\alpha\}$ is a set of coordinates on the brane and embedding in target space is given by $X^\mu(\sigma^\alpha)$. Most computations are done in static gauge: $X^\alpha = \sigma^\alpha$, $X^i = X^i(\sigma^\alpha)$. Massless closed string fields $G_{\mu\nu}, B_{\mu\nu}$ and $\Phi$ are functions of $X^\mu$ and massless open string field $A^\alpha$ is a function of $\sigma^\alpha$. The action (1) describes the coupling of a brane to NS-NS background bulk fields $G_{\mu\nu}$, $B_{\mu\nu}$ and $\Phi$. Tilde ˜ denotes the induced quantities: $\tilde{G}_{\alpha\beta} = G_{\mu\nu}\frac{\partial X^\mu}{\partial\sigma^\alpha}\frac{\partial X^\nu}{\partial\sigma^\beta}$, $\tilde{B}_{\alpha\beta} = B_{\mu\nu}\frac{\partial X^\mu}{\partial\sigma^\alpha}\frac{\partial X^\nu}{\partial\sigma^\beta}$. The action for the bulk massless fields is also well known. For example in bosonic case it is given to the leading order (in Einstein frame) by

$$S_{bulk} = \frac{1}{2k^2} \int d^D x \sqrt{-G}(R - \frac{1}{12}e^{2\gamma\Phi}H_{\mu\nu\rho}H^{\mu\nu\rho} + \gamma\partial_\mu\Phi\partial^\mu\Phi) \quad (2)$$

Where $H_{\mu\nu\rho} = 3\partial_{[\mu}B_{\nu\rho]}$. Subleading terms can be found in [5].

There are different ways to determine brane and bulk actions. One of them is to compute string tree amplitudes for massless fields, expand them in powers of $\alpha'$ and look for terms in the effective action to reproduce them. In this way the $\alpha'^2$ curvature corrections to the brane action in superstring case [6] and $\alpha'$ curvature corrections to the brane action for bosonic string [7] were found. Another way is to compute renormalization group beta function for the field theory of strings on the world-sheet. Consistency condition of (super)conformal invariance impose that these beta-functions should vanish. Treating massless string fields as background we get equations of motion from which we can derive the effective action. It is believed that the the two approaches are perturbatively equivalent (see for example [8],[5]). In this paper we perform another check of the correspondence and also compute certain $\alpha'$ corrections to D-brane action evaluating the two-loop beta-function in the sigma model.

---
[1]For a review see [4]

In section 2 we present the $\sigma$-model relevant to the case of a single D-brane. Using the background field method we compute one-loop beta function for the scalar fields $\Phi_i$ describing transverse fluctuations of the brane in a curved ambient space. In section 3 we reproduce some of the results of [7] in the $\sigma$-model language. In particular we determine the coefficients of the possible $O(\alpha')$ curvature terms in the effective action and discuss field redefinitions.

The case of a stack of N coincident D-branes is somewhat more complicated. Now we expect the effective action of the brane to be generalized to a non-abelian theory with gauge group U(N). In this case fields get promoted to U(N) matrices. One of the open problems is how to expand the square root of the determinant since the ordering of the non-commutative fields is not clear. We are not addressing this problem here. In section 4 we describe sigma model with boundary fermions and compute one-loop beta function for the non-abelian field $A_\alpha$ in curved background. Setting beta-function to zero gives equations of motion consistent with DBI action. It is interesting to study the higher order corrections to DBI action. In non-abelian case in curved background there is a new class of terms that could appear. In section 5 we compute corrections of the form $R_{ijkl}[\Phi^i\Phi^j][\Phi^k\Phi^l]$ to the effective action in bosonic and superstring case. We discuss the results in section 6.

## 2  $\sigma$-model actions

The $\sigma$-model action for a single Dp-brane in bosonic string theory contains 2 terms-bulk and boundary:

$$S_\Sigma = \frac{1}{4\pi\alpha'}\int_\Sigma d^2\sigma\sqrt{g}[g^{ab}G_{\mu\nu}(X)\partial_a X^\mu \partial_b X^\nu + i\epsilon^{ab}B_{\mu\nu}(X)\partial_a X^\mu \partial_b X^\nu + \alpha' R\Phi(X)] \tag{3}$$

$$S_{\partial\Sigma} = \int d\theta[-i\partial_\theta \sigma^\alpha A_\alpha + \frac{1}{2\pi\alpha'}\partial_r X^\mu \Phi_\mu] \tag{4}$$

Here fields $G_{\mu\nu},B_{\mu\nu}$ and $\Phi$ are functions of $X^\mu$ and $A_\alpha$ and $\Phi_\mu$ are functions of $\sigma^\alpha$. Consider for the time being only terms involving $G_{\mu\nu}$ and $\Phi_i$. They can also be rewritten as

$$S = \frac{1}{2\pi\alpha'}\int_\Sigma d^2 z G_{\mu\nu}\partial X^\mu \bar\partial X^\nu + \frac{1}{2\pi\alpha'}\int_{\partial\Sigma} d\theta \partial_r X^i \Phi_i \tag{5}$$

The last term was obtained using T-duality from $\sigma$-model of the open string theory without D-branes. We are going to use background field method expanding this action near a bare classical solution $X^\mu = \tilde X^\mu + \pi^\mu$. It is easier to use expansion in normal coordinates $\xi^\mu$ in space-time and $\zeta^\alpha$ on the brane following [9],[2]. For our purposes it is sufficient to expand to the fourth order in $\xi$ and $\zeta$:

$$S_\Sigma[X] = \frac{1}{2\pi\alpha'}\int d^2z\{G_{\mu\nu}\partial\tilde{X}^\mu\bar{\partial}\tilde{X}^\nu + G_{\mu\nu}(\partial\tilde{X}^\mu\bar{\nabla}\xi^\nu + \bar{\partial}\tilde{X}^\mu\nabla\xi^\nu)$$
$$+ R_{\mu\nu\rho\sigma}\partial\tilde{X}^\mu\bar{\partial}\tilde{X}^\sigma\xi^\nu\xi^\rho + G_{\mu\nu}\nabla\xi^\mu\bar{\nabla}\xi^\nu + \frac{1}{3}D_\mu R_{\nu\rho\sigma\tau}\partial\tilde{X}^\nu\bar{\partial}\tilde{X}^\tau\xi^\mu\xi^\rho\xi^\sigma$$
$$+ \frac{2}{3}R_{\mu\nu\rho\sigma}(\partial\tilde{X}^\mu\bar{\nabla}\xi^\sigma + \bar{\partial}\tilde{X}^\mu\nabla\xi^\sigma)\xi^\nu\xi^\rho + \frac{1}{3}R_{\mu\nu\rho\sigma}\nabla\xi^\mu\bar{\nabla}\xi^\sigma\xi^\nu\xi^\rho$$
$$+ \frac{1}{12}(D_\mu D_\nu R_{\rho\sigma\tau\omega} + 4R^{\mu'}_{\mu\nu\rho}R_{\mu'\sigma\tau\omega})\partial\tilde{X}^\rho\bar{\partial}\tilde{X}^\omega\xi^\mu\xi^\nu\xi^\sigma\xi^\tau$$
$$+ \frac{1}{4}D_\mu R_{\nu\rho\sigma\tau}(\partial\tilde{X}^\nu\bar{\nabla}\xi^\tau + \bar{\partial}\tilde{X}^\nu\nabla\xi^\tau)\xi^\mu\xi^\rho\xi^\sigma\} + \ldots \quad (6)$$

$$S_{\partial\Sigma}[X] = \frac{1}{2\pi\alpha'}\int d\theta\big[\partial_r\tilde{X}^i + \nabla_r\xi^i + \partial_r\tilde{X}^\lambda(\frac{1}{3}R^i_{\mu\nu\lambda}\xi^\mu\xi^\nu + \frac{1}{12}D_\mu R^i_{\nu\rho\lambda}\xi^\mu\xi^\nu\xi^\rho$$
$$+ (\frac{1}{60}D_\mu D_\nu R^i_{\rho\sigma\lambda} - \frac{1}{45}R^i_{\mu\nu\tau}R^\tau_{\rho\sigma\lambda})\xi^\mu\xi^\nu\xi^\rho\xi^\sigma)\big]\big[\Phi_i + \tilde{D}_\alpha\Phi_i\zeta^\alpha$$
$$+ \frac{1}{2}\tilde{D}_\alpha\tilde{D}_\beta\Phi^j\zeta^\alpha\zeta^\beta + \frac{1}{6}\tilde{D}_\alpha\tilde{D}_\beta\tilde{D}_\gamma\Phi^j\zeta^\alpha\zeta^\beta\zeta^\gamma$$
$$+ \frac{1}{24}\tilde{D}_\alpha\tilde{D}_\beta\tilde{D}_\gamma\tilde{D}_\delta\Phi^j\zeta^\alpha\zeta^\beta\zeta^\gamma\zeta^\delta\big] + \ldots \quad (7)$$

Here $D_\mu$ is the usual covariant derivative with Levi-Civita connection, $\tilde{D}_\alpha$ is the covariant derivative on the brane constructed using the induced metric and

$$\nabla\xi^\mu = \partial\xi^\mu + \Gamma^\mu_{\nu\rho}\xi^\nu\partial\tilde{X}^\rho, \bar{\nabla}\xi^\mu = \bar{\partial}\xi^\mu + \Gamma^\mu_{\nu\rho}\xi^\nu\bar{\partial}\tilde{X}^\rho \text{ and } \nabla_r\xi^\mu = \partial_r\xi^\mu + \Gamma^\mu_{\nu\rho}\xi^\nu\partial_r\tilde{X}^\rho$$

Using equations of motion we obtain additional boundary term:

$$\frac{1}{2\pi\alpha'}\int d^2z G_{\mu\nu}(\partial X^\mu\bar{\nabla}\xi^\nu + \bar{\partial}X^\mu\nabla\xi^\nu) = \frac{1}{2\pi\alpha'}\int d\theta\partial_r X^\mu G_{\mu\nu}\xi^\nu \quad (8)$$

In the bulk action kinetic term is multiplied by $G_{\mu\nu}(\tilde{X})$. One way of bringing it to the standard form is introducing vielbein field $V^A_\mu$, $G_{\mu\nu} = V^A_\mu V^B_\nu \eta_{AB}$ and switching to the tangent space quantities. Here we choose another approach. We expand the metric as $G_{\mu\nu} = \eta_{\mu\nu} + 2kH_{\mu\nu}$ and expand sigma-model and effective actions in powers of $H$.

The "Neumann" and "Dirichlet" propagators on the disc are given correspondingly by:

$$N^{\alpha\beta}(z,z') = <\xi^\alpha(z)\xi^\beta(z')> = \frac{\alpha'}{2}\eta^{\alpha\beta}\{-\ln|z-z'|^2 - \ln|1-\bar{z}z'|^2\} \quad (9)$$

$$D^{ij}(z,z') = <\xi^i(z)\xi^j(z')> = \frac{\alpha'}{2}\delta^{ij}\{-\ln|z-z'|^2 + \ln|1-\bar{z}z'|^2\} \quad (10)$$

The superstring sigma model contains additional pieces with fermions. The supersymmetric extension of part of the bulk action containing $G_{\mu\nu}$ field is given by

$$S_\Sigma = \frac{1}{2\pi\alpha'} \int_\Sigma d^2z \{G_{\mu\nu}(\partial\tilde{X}^\mu\bar{\partial}\tilde{X}^\nu + \frac{\alpha'}{2}\Psi^\mu\bar{\nabla}\Psi^\nu + \frac{\alpha'}{2}\bar{\Psi}^\mu\nabla\bar{\Psi}^\nu) + \frac{\alpha'^2}{2}R_{\mu\nu\rho\sigma}\Psi^\mu\Psi^\nu\bar{\Psi}^\rho\bar{\Psi}^\sigma\} \qquad (11)$$

The boundary action is obtained by T-duality from the supersymmetric version of Wilson loop. The pieces containing $\Phi_i$ fields are given by

$$S_{\partial\Sigma} = \frac{1}{2\pi\alpha'} \int_{\partial\Sigma} d\theta \big[\partial_r\tilde{X}^i\Phi_i + \alpha'(\psi^\alpha\bar{\psi}^i - \psi^i\bar{\psi}^\alpha)\partial_\alpha\Phi_i\big] \qquad (12)$$

Here $\psi^\mu = \Psi^\mu|_{\partial\Sigma}$ We will not be interested in the supersymmetric extension of the term with $B_{\mu\nu}$ field.

Expansion of the additional fermionic terms in normal coordinats sufficient for two-loop computation is:

$$\frac{1}{4\pi} \int_\Sigma dz^2 \{G_{\mu\nu}(\Psi^\mu\bar{\nabla}\Psi^\nu + \bar{\Psi}^\mu\nabla\bar{\Psi}^\nu) + R_{\mu\nu\rho\sigma}(\frac{1}{3}\xi^\nu\xi^\rho(\Psi^\mu\bar{\nabla}\Psi^\sigma + \bar{\Psi}^\mu\nabla\bar{\Psi}^\sigma) + \alpha'\Psi^\mu\Psi^\nu\bar{\Psi}^\rho\bar{\Psi}^\sigma - \frac{1}{2}\bar{\partial}\tilde{X}^\mu\xi^\nu\Psi^\rho\Psi^\sigma - \frac{1}{2}\partial\tilde{X}^\mu\xi^\nu\bar{\Psi}^\rho\bar{\Psi}^\sigma)\} \quad (13)$$

and

$$\frac{1}{2\pi\alpha'} \int_{\partial\Sigma} d\theta \big[\alpha'(\psi^\alpha\bar{\psi}^i - \psi^i\bar{\psi}^\alpha)(\partial_\alpha\Phi_i + \tilde{D}_\beta\partial_\alpha\Phi_i\zeta^\beta + \frac{1}{6}(3\tilde{D}_\beta\tilde{D}_\gamma\partial_\alpha\Phi_i + \tilde{R}^\delta_{\beta\gamma\alpha}\partial_\delta\Phi_i)\zeta^\beta\zeta^\gamma) \quad (14)$$

The fermionic propagators are

$$<\Psi^\mu(z)\Psi^\nu(w)> = \frac{\eta^{\mu\nu}}{z-w}, <\Psi^\mu(z)\bar{\Psi}^\nu(w)> = \frac{i\eta^{\mu\nu}}{1-z\bar{w}} \qquad (15)$$

Our strategy will be to compute renormalization group beta-function corresponding to the coupling $\frac{1}{2\pi\alpha'}\int d\theta\partial_r X^i\Phi_i$. Following [11] we define beta-function for the field $\Phi_i$ to be:

$$\beta_i^\phi(\Phi^{bare}) = -\frac{d}{d\ln\Lambda}\Phi_i^{bare}(\Phi) \qquad (16)$$

Here

$$\Phi_i^{bare}(\Phi) = \Phi_i + \sum_n K_i^{(n)}(\Phi)(\ln\Lambda)^n \qquad (17)$$

Solving for $\Phi_i$ one can obtain $\beta_i^\phi(\Phi^{bare})$ in terms of $\Phi_i^{bare}$. Now explicit dependence on the $\ln \Lambda$ must cancel which is a good test of calculations. Setting the beta-function to zero will give us equations on $G_{\mu\nu}$ and $\Phi_i$.

Let us compute one-loop beta-function to the first order in $H_{\mu\nu}$ and $\Phi_i$ in bosonic case. For simplicity consider the case when $H_{\alpha i} = 0$, $H_{\mu\nu}$ depends only on $\sigma^\alpha$ and $X^i = const$. The general procedure is to leave normal coordinates $\partial_r \xi^i$ and $\xi^\alpha$ in the boundary action unchanged and express normal coordinates $\xi^i$ (without derivative $\partial_r$) and $\zeta^\alpha$ in the boundary action in terms of $\xi^\alpha$. Under the above conditions this means: $\xi^i|_{\partial\Sigma} = 0$ (but not $\partial_r \xi^i$) and $\xi^\alpha|_{\partial\Sigma} = \zeta^\alpha$. We have to evaluate the following set of diagrams:

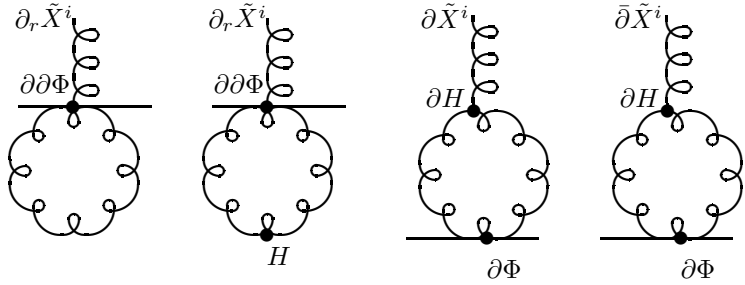

Figure 1. *One-loop corrections to the boundary vertex. Here horizontal lines represent the boundary and wiggly lines correspond to bosonic string coordinates.*

The last two diagrams with bulk vertices combine to produce contribution proportional to $\partial_r \tilde{X}^i$. Their computation involves integration over $d^2z$ and no integration over $\theta$. Bulk integration can be done as follows [10]:

$$\int d^2z = \int r dr \int d\theta = \int r dr \oint \frac{dw}{iw} \qquad (18)$$

where $w = e^{i\theta}$. First one needs to perform contour integration and then integration over $r$ introducing a cutoff $\Lambda$ (so that $N^{\alpha\beta}(u,u) = -\alpha' \eta^{\alpha\beta} \ln \Lambda^2$ for the point $u$ on the boundary of the disc). As a result we get:

$$\beta^\Phi{}_i = \alpha' \partial^2 \Phi_i + 2k\alpha' \left( -H_{\alpha\beta} \partial^\alpha \partial^\beta \Phi_i - \partial^\alpha H_{\alpha\beta} \partial^\beta \Phi_i + \frac{1}{2} \partial^\alpha H^\beta{}_\beta \partial_\alpha \Phi_i + \partial^\alpha H_{ij} \partial_\alpha \Phi_j \right) \qquad (19)$$

It is easy to see that at this order beta-function is proportional to the equations of motion coming from:

$$S_{eff} = -\tau_p \int d^{p+1}\sigma \sqrt{-\tilde{G}} \qquad (20)$$

where for induced metric on the D-brane we have

$$\tilde{G}_{\alpha\beta}(\sigma^\alpha, X^i + \tilde{\Phi}^i) = \eta_{\alpha\beta} + 2kH_{\alpha\beta} + \partial_\alpha \tilde{\Phi}^i \partial_\beta \tilde{\Phi}^i + 2kH_{ij} \partial_\alpha \tilde{\Phi}^i \partial_\beta \tilde{\Phi}^j \qquad (21)$$

if we identify $\Phi_i \equiv \tilde{\Phi}^i$. More precisely we get:

$$\delta S_{eff} = \frac{\tau_p}{\alpha'} \int d^{p+1}\sigma \sqrt{-\tilde{G}} (\beta^\Phi(\tilde{\Phi}))^i G_{ij} \delta\tilde{\Phi}^j \qquad (22)$$

Computations in more general cases are also straightforward. For example if we relax condition $H_{i\alpha} = 0$ and allow $H_{\mu\nu}$ depend on $X^i$ then boundary term without $\Phi_i$ will also contribute to the beta-function and we have

$$\xi^i|_{\partial\Sigma} = \tfrac{1}{2}\Gamma^i_{\alpha\beta}\xi^\alpha\xi^\beta + \tfrac{1}{6}\Gamma^i_{\alpha\beta\gamma}\xi^\alpha\xi^\beta\xi^\gamma + \tfrac{1}{24}\Gamma^i_{\alpha\beta\gamma\delta}\xi^\alpha\xi^\beta\xi^\gamma\xi^\delta + O(\xi^5)$$

Computations to higher orders in $H_{\mu\nu}$ and for $X^i(\sigma^\alpha) \neq const$ become more tedious.

## 3 Curvature corrections to the action of a single D-brane

In [7] the authors were able to determine first order curvature corrections to Born-Infeld action in the case of bosonic string analyzing tree level amplitudes corresponding to scattering of massless closed string fields off the brane. At order $O((\alpha')^0)$ they showed agreement between string and field theory amplitudes using the expansion of bulk and brane actions (2) and (1). Massless closed and open string fields are redefined as:

$G_{\mu\nu} = \eta_{\mu\nu} + 2kH_{\mu\nu}$, $\Phi = k\sqrt{\frac{D-2}{4}}\phi$, $B_{\mu\nu} = -2kb_{\mu\nu}$,

$\tilde{\Phi}^i = \frac{1}{\sqrt{\tau_p}}\lambda^i$ and $A_\alpha = \frac{1}{2\pi\alpha'\sqrt{\tau_p}}a_\alpha$

Comparison of the amplitudes on string and field theory side fixes the normalization constant of string amplitudes in terms of $\tau_p$ and $k$. At order $O(\alpha')$ there are five possible terms that can contribute to graviton scattering:

$$S^{(1)}_{brane} = -\frac{1}{2k_p{}^2} \int d^{p+1}\sigma \sqrt{-\tilde{G}} \{\beta_0 \tilde{R} + \beta_1 K^i_{\alpha\beta} K^{\alpha\beta}_i + \beta_2 K^{i\alpha}_\alpha K^\beta_{i\beta} + $$
$$\beta_3 R_{\mu\nu} \perp^{\mu\nu} + \beta_4 R_{\mu\nu\rho\sigma} \perp^{\mu\rho} \perp^{\nu\sigma} \} \qquad (23)$$

For the geometry of submanifold we closely follow [7]. Denote $n^\mu_i$ some orthonormal basis of normal vectors to the submanifold $\Sigma$ representing the embedded p-brane. One can define the projection operator

$$\perp^{\mu\nu} = \sum_{i=p+1}^{D-1} n^\mu_i n^\nu_i = G^{\mu\nu} - \tilde{G}^{\mu\nu} \qquad (24)$$

Where,

$$\tilde{G}^{\mu\nu} = \frac{\partial X^\mu}{\partial \sigma^\alpha}\frac{\partial X^\nu}{\partial \sigma^\beta}\tilde{G}^{\alpha\beta} \qquad (25)$$

We need to know the expressions for the 5 terms in the action (23). $R_{\mu\nu\rho\sigma}$, $R_{\mu\nu}$ are the usual Riemann and Ricci tensors. $\tilde{R}$ is the scalar curvature computed using the induced metric. $K^i_{\alpha\beta}$ is the second fundamental form defined by

$$K^i_{\alpha\beta} = \Big(\frac{\partial^2 X^\mu}{\partial\sigma^\alpha \partial\sigma^\beta} + \frac{\partial X^\nu}{\partial\sigma^\alpha}\frac{\partial X^\rho}{\partial\sigma^\beta}\Gamma^\mu_{\nu\rho}\Big)n^i_\mu \qquad (26)$$

One also needs bulk action to the $O(\alpha')$ order. In [7] $S^{(1)}_{R^2 bulk}$ was chosen in the Gauss-Bonnet form:

$$S^{(1)}_{R^2 bulk} = \frac{1}{2k^2}\int d^D x \frac{\alpha'}{4} e^{\gamma\Phi}\big(R^{\mu\nu\rho\sigma}R_{\mu\nu\rho\sigma} - 4R^{\mu\nu}R_{\mu\nu} + R^2\big) \qquad (27)$$

Note that each of the above actions (23) and (27) is affected by field redefinitions [7]. As a result, the coefficients $\beta_0,\beta_1,\beta_2,\beta_3$ and $\beta_4$ are dependent on two parameters. Comparison of the amplitudes determines the coefficients:

$\beta_0 = 1+\alpha$, $\beta_1 = -1+\alpha$, $\beta_2 = 1-\alpha$, $\beta_3 = \beta$, $\beta_4 = -\alpha$ and $\frac{1}{2k_p} = \tau_p$

The choice of $S^{(1)}_{R^2 bulk}$ in the Gauss-Bonnet form implies that free parameters $\alpha$ and $\beta$ are actually fixed: $\alpha = \beta = 0$

Let us see how this result can be re-derived from the $\sigma$-model prospective. Since this is a two-loop computation, we also need to take into account one-loop beta-function for $G$-field as well as one-loop equations of motion. At one loop:

$$\beta^{G(1)}_{\mu\nu} = -\alpha' R_{\mu\nu} + O(\alpha') \qquad (28)$$

To simplify the computation we may assume that $H_{\mu\nu}$ depends only on $\sigma^\alpha$ and $H_{i\alpha} = 0$. As in the previous section expand all five terms from (23) and $\sigma$-model actions (6) and (7) to the linear order in $H_{\mu\nu}$ and quadratic order in $\tilde{\Phi}^i$. It turns out that only one two-loop diagram gives contribution proportional to $\ln\Lambda$ :

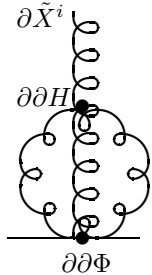

Figure 2. *Two-loop diagram with non-zero $\ln\Lambda$ contribution.*

It has the following structure: $\partial^\alpha \partial^\beta H_{ij} \partial_\alpha \partial_\beta \tilde{\Phi}^j$. This means that no other contractions between derivatives of $H_{\mu\nu}$ and derivatives of $\tilde{\Phi}^i$ modulo $R_{\mu\nu}$ could appear as a result of variation of (23). Performing the variation of (23) with respect to $\tilde{\Phi}^i$ explicitly gives an over-determined system of equations on the coefficients $\beta_0$, $\beta_1$, $\beta_2$, $\beta_3$ and $\beta_4$ which reduces to: $\beta_1 = -\beta_2 = 0$, $\beta_0 = -2\beta_4$ and $\beta_3$ is undetermined. Computing the diagram on Figure 2 fixes the normalization. All together we find: $\beta_0 = 2$, $\beta_1 = \beta_2 = 0$, $\beta_3$-undetermined, $\beta_4 = -1$ and $\frac{1}{2k_p} = \tau_p$. This corresponds to the choice of $\alpha = 1$ and is in agreement with the result in [7]. As was pointed out in [5] field redefinitions on the effective action side correspond to the different choices of the renormalization schemes for $\sigma$-model computations. Thus in our scheme we would not get the $O(\alpha')$ corrections to the bulk action in the Gauss-Bonnet form. A non-trivial check of the above computation is the cancellation of all $\ln \Lambda$ terms in the beta-function.

What happens in the case of the superstring? It is easy to check that diagrams containing fermion lines don't contribute at one loop. Thus $O((\alpha')^0)$ part of the brane action is not affected. At two loops only one diagram (Figure 3) gives contribution proportional to $\ln \Lambda$. In fact it precisely cancels the contribution of the corresponding bosonic diagram (Figure 2). Thus all coefficients $\beta_0$, $\beta_1$, $\beta_2$ and $\beta_4$ are zero. This means that there are no corrections to the BI action linear in curvature in the superstring case.

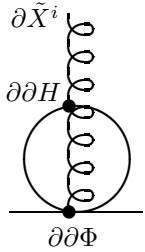

Figure 3. *Two-loop diagram with fermions with non-zero $\ln \Lambda$ contribution.*

## 4 $\sigma$-model analysis of a system of coincident D-branes

$\sigma$-model analysis of a stack of coincident Dp-branes is more complicated. Instead of using $S_{\partial \Sigma}$ (7) we should rather be using a gauge invariant Wilson loop

$$\tilde{S}_{\partial \Sigma} = -\log tr P \exp\{\oint_{\partial \Sigma} d\theta (iA_\alpha \partial_\theta \sigma^\alpha - \frac{1}{2\pi\alpha'} \Phi_i \partial_r X^i)\} \tag{29}$$

Computations using the expansion of Wilson loops $trP \exp(i \int_{\partial \Sigma} A_\mu dX^\mu)$ are straightforward (see for example [11],[12]). There is however a way to avoid it introducing the auxiliary boundary fermions [13, 10]. Now Wilson loop can be rewritten as a functional integral over the boundary fermions coupled to the fields $A_\alpha$ and $\Phi_i$ via

$$e^{-S_{\partial\Sigma}(A,\Phi)} = \sum_{k=1}^{N}[d\lambda^\dagger d\lambda]e^{i\frac{2\pi k}{N}[\lambda^\dagger\lambda(\tau=\tau_0)+\frac{N}{2}-1]}e^{-\int_0^{2\pi}d\theta\lambda^\dagger(\frac{d}{d\theta}+\frac{1}{2\pi\alpha'}\partial_r X^i\Phi_i-i\partial_\theta\sigma^\alpha A_\alpha)\lambda}$$
(30)

Thus we have new auxiliary fermionic fields with propagators

$$<\lambda_a(\theta)\lambda_b^\dagger(\theta')> = \frac{1}{2}\delta_{ab}sign(\theta-\theta')$$
(31)

Boundary fermions couple to the string coordinates $X^\mu$ via $N \times N$ Hermitian traceless matrices $A_\alpha$ and $\Phi_i$ viewed as background fields.

In [10] this model was used to study world-volume potentials on a stack of coincident D-branes and world-volume couplings of NS fluxes which are responsible for Myers' dielectric effect [14]. In this section we want to study the effect of introduction of non-trivial embedding of a stack of coincident D-branes in a curved target space.

As a first example let us re-derive some of the results of [11] in this model as opposed to the expansion of the Wilson loop. For simplicity consider the case of Neumann boundary condition in all directions. One-loop beta-function for the non-abelian gauge field $A_\alpha$ was shown to be proportional to $(D^{A+\Gamma})^\beta F_{\alpha\beta}$. $D^{A+\Gamma}$ is a covariant derivative constructed using gauge field and Levi-Civita connection. This is simply the equations for the gauge field $A_\alpha$ in the background gravitational field. In order to compute beta-function for $A_\alpha$ we need to consider renormalization of the coupling

$$-i\int d\theta \lambda_a^\dagger (A_\alpha)^{ab}\lambda_b \partial_\theta X^\alpha$$
(32)

Expansion of this term is similar to (7) since the boundary fermions $\lambda$ are quantum fields from the beginning. To the first order in $H_{\alpha\beta}$ we get

$$-i\int d\theta\lambda^\dagger[\partial_\theta\tilde{X}^\alpha(A_\alpha+\partial_\beta A_\alpha\xi^\beta+\frac{1}{2}(\partial_\beta\partial_\gamma A_\alpha-\Gamma_{\beta\gamma}^\delta\partial_\delta A_\alpha-\partial_\alpha\Gamma_{\beta\gamma}^\delta A_\delta)\xi^\beta\xi^\gamma)$$
$$+\partial_\theta\xi^\alpha(A_\alpha+(\partial_\beta A_\alpha-\Gamma_{\alpha\beta}^\gamma A_\gamma)\xi^\beta)]\lambda \quad (33)$$

However in order to make comparison easier it is better to write it in a form:

$$-i\int d\theta\{\lambda_a^\dagger(\partial_\theta\tilde{X}^\alpha A_\alpha - \partial_\theta\tilde{X}^\alpha\xi^\beta\tilde{F}_{\alpha\beta} - \frac{1}{2}\partial_\theta\tilde{X}^\alpha\xi^\beta\xi^\gamma\partial_\gamma\tilde{F}_{\alpha\beta} -$$
$$\frac{1}{2}\partial_\theta\xi^\alpha\xi^\beta\tilde{F}_{\alpha\beta} + \frac{1}{2}\partial_\theta\tilde{X}^\alpha\xi^\gamma\xi^\delta\Gamma_{\gamma\delta}^\beta\tilde{F}_{\alpha\beta})^{ab}\lambda_b -$$
$$(A_\alpha\xi^\alpha + \frac{1}{2}\partial_\beta A_\alpha\xi^\alpha\xi^\beta - \frac{1}{2}\Gamma_{\beta\gamma}^\alpha A_\alpha\xi^\beta\xi^\gamma)^{ab}\partial_\theta(\lambda_a^\dagger\lambda_b)\} \quad (34)$$

One can get it using integration by parts along the lines of [15] (The authors of [15] considered abelian gauge field in a flat background). Here $\tilde{F}_{\alpha\beta} =$ $\partial_\alpha A_\beta - \partial_\beta A_\alpha$. Now it is easy to see which diagrams will contribute to the corresponding terms in $D^{(A+\Gamma)}F$. For example, consider the following diagram involving bulk vertex:

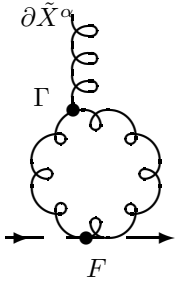

Figure 4. *Example of one-loop diagram with boundary fermions (represented by dashed lines).*

Together with the diagram with complex conjugate bulk vertex it combines to give: $-i\alpha' \Gamma_{\beta,\alpha\gamma} \partial_\theta \tilde{X}^\alpha F^{\beta\gamma} \ln \Lambda$. At this order:

$$\delta\left(-\tau_p Tr \int d^{p+1}\sigma \sqrt{-det(\tilde{G}+F)}\right) = \frac{\tau_p}{\alpha'} \int d^{p+1}\sigma \sqrt{-\tilde{G}} \beta^A{}_\alpha \delta A^\alpha \qquad (35)$$

When there are several vertices on the boundary one may worry that there will be many regions of integration corresponding to relative positions of the angles on the boundary. However, as was shown in [10] using the symmetry properties of the $\xi$ propagators on the boundary and the fact that they are double periodic functions of angles it is possible to significantly reduce the number of regions. The perturbation theory becomes path-ordered like in ordinary quantum mechanics. In this case the positions of ordered vertices lie in the interval $[\theta_i, \theta_f] \subset [0, 2\pi]$.

## 5 $R\Phi^4$ corrections to the effective action

It is clear that this model provides a simple way of computing both world-volume potentials and derivative corrections to the brane action. The $O(\alpha')$ corrections to the effective action (23) have overall trace in the non-abelian case. Otherwise the analysis of section 3 holds in this case as well since it involvs terms quadratic in $\Phi_i$. Let us study another possible class of terms that could appear in the effective action: $R\Phi^4$. Consider two of them:

$$\int d\sigma^{p+1} Tr(aR_{ijkl}[\tilde{\Phi}^i\tilde{\Phi}^j][\tilde{\Phi}^k\tilde{\Phi}^l] + bR_{ijkl}\tilde{\Phi}^i\tilde{\Phi}^k\tilde{\Phi}^j\tilde{\Phi}^l) \qquad (36)$$

Using the cyclic property of trace and the symmetries of $R_{ijkl}$ it is easy to see that for $R_{ijkl}$ independent of $\tilde{\Phi}^i$ the second term is zero. Thus $b$ cannot be determined at two loops. In order to determine the first coefficient we have to evaluate certain two-loop diagrams. In bosonic case at order $O(H)$ the relevant diagram is:

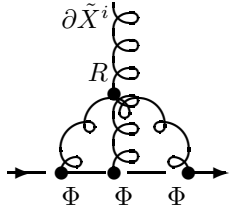

Figure 5. *Two-loop diagram contributing to $TrR_{ijkl}[\tilde{\Phi}^i\tilde{\Phi}^j][\tilde{\Phi}^k\tilde{\Phi}^l]$*

To compute the diagram we will have to evaluate the following integral for different orderings of $\{\theta_1, \theta_2, \theta_3\}$:

$$\int \frac{d^2z d\theta_1 d\theta_2 d\theta_3 (1-r^2)^2}{u_3(z-u_1)(\bar{z}-\frac{1}{u_1})(z-u_2)(\bar{z}-\frac{1}{u_2})(\bar{z}-\frac{1}{u_3})^2} \tag{37}$$

Here $u_i = e^{i\theta_i}$. We rewrite $\int d^2z = \int r dr \oint \frac{dw}{iw}$, $z = rw$ and $\bar{z} = \frac{r}{w}$. First integrations over $\theta_1$ and $\theta_3$ are performed, then contour integration over $w$, and finally integration over $r$ introducing the cutoff. One integration over $\theta$ is left out since we want to find the renormalization of the $\frac{1}{2\pi\alpha'}\int d\theta \partial_r \tilde{X}^i \Phi_i$ coupling. A somewhat lengthy but straightforward computation leads to the following answer for the sum of this diagram and diagram with complex conjugate bulk vertex:

$$\frac{-1}{(2\pi\alpha')^3}\frac{\alpha'^2}{4}(-\ln\Lambda)\int d\theta \partial\tilde{X}^i R_{ijkl}[\Phi^j[\Phi^k\Phi^l]] \tag{38}$$

Interpreting (38) as equations of motion, we find: $a = -\frac{\alpha'\tau_p}{16(2\pi\alpha')^2}$.

In the supersymmetric case there is additional term:

$$\int_{\partial\Sigma} d\theta(-\alpha'\psi^i\psi^j[\Phi_i\Phi_j]) \tag{39}$$

The diagram on Figure 6 is the only two-loop diagram involving fermions proportional to $R\Phi^3$.

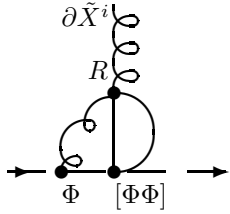

Figure 6. *Two-loop diagram with fermions contributing to* $TrR_{ijkl}[\Phi^i\Phi^j][\Phi^k\Phi^l]$

It cancels the contribution of the corresponding bosonic diagram. Thus, $R\Phi^4$ corrections are absent in this case.

## 6 Discussion

In this paper we used two-loop sigma model computation to determine certain gravitational corrections to D-brane action. In bosonic case the $\alpha'$ corrections (23) were in agreement with those found in [7]. Corrections depending on dilaton and $B$-field is the subject of the future work. Interesting results in this direction using different techniques were obtained in [16]. In superstring case we analyzed the possibility of $(\alpha')^2$ corrections of the form (36). Those terms could be of interest in AdS/CFT correspondence or dynamics of giant gravitons.

For example in the case of spaces of constant curvature with curvature independent of the transverse coordinates (36) is proportional to $Tr([\Phi^i, \Phi^j]^2)$ which is important for Myers' dielectric effect [14]. Thus it could be considered as a next order correction since the effective expansion parameter in the sigma model is $\frac{\alpha'}{R_c^2}$. However in the present paper we showed that those corrections are absent in the superstring case.

### Acknowledgments

I would like to thank Tomasz Taylor for bringing my attention to this problem and for illuminating discussions. Also I wish to thank Ahmad Ghodsi and Oleg Andreev for correspondence. This work was supported in part by NSF grant PHY-99-01057